# Positive Contrast Susceptibility MR Imaging Using GPU-based Primal-Dual Algorithm

Haifeng Wang, Fang Cai, Caiyun Shi, Jing Cheng, Shi Su, Zhilang Qiu, Guoxi Xie, Hanwei Chen, Xin Liu and Dong Liang, *Senior Member, IEEE*

*Abstract*—The susceptibility-based positive contrast MR technique was applied to estimate arbitrary magnetic susceptibility distributions of the metallic devices using a kernel deconvolution algorithm with a regularized L-1 minimization. Previously, the first-order primal-dual (PD) algorithm could provide a faster reconstruction time to solve the L-1 minimization, compared with other methods. Here, we propose to accelerate the PD algorithm of the positive contrast image using the multi-core multi-thread feature of graphics processor units (GPUs). The some experimental results showed that the GPU-based PD algorithm could achieve comparable accuracy of the metallic interventional devices in positive contrast imaging with less computational time. And the GPU-based PD approach was 4~15 times faster than the previous CPU-based scheme.

*Clinical Relevance*—This can estimate arbitrary magnetic susceptibility distributions of the metallic devices with the processing efficacy of 4~15 times faster than before.

I. INTRODUCTION

Today, MR-compatible metallic medical devices, such as brachytherapy seeds and biopsy needles, have been widely used in many clinical applications [1, 2]. However, these interventional metallic devices have exhibited black hole phenomena in conventional magnetic resonance images, because of the negative contrast [3]. The susceptibility-based positive contrast MRI is a technique that combines modified pulse sequences and post-processing to achieve positive contrast images of MR-compatible metal devices. It requires the phase generated by the difference of the magnetic susceptibility of the tissue itself (local magnetic field effect) by collecting the spin echo data with a short time offset in the pulse direction and subtracting the spin echo data without offset in the readout direction. After that, they can be combined with the quantitative magnetic susceptibility image reconstruction algorithm and implemented to organize the intensity of magnetic susceptibility to contrast MRI. However, the quantitative susceptibility image reconstruction of the above process is usually an L-1 minimization problem with regularization constraints, and an accurate approximate solution needs to continually be solved iteratively. Because of the multiple slice data for 3D image reconstruction and the multi-coil acquisition data from parallel imaging, the entire reconstruction calculation process often takes several minutes.

Currently, the existing schemes of positive contrast MRI recon are based on the traditional central processing unit (CPU), to implement a sequential iterative solution method, such as the nonlinear conjugate gradient algorithm (NLCG) [4], the fast iterative soft-thresholding algorithm (FISTA) [5], alternating direction method of multipliers (ADMM) [6], and first-order primal-dual algorithm (PD) [7]. The previous experiments have evaluated the performance of the PD algorithm [8] better than NLCG [4], FISTA [5], and ADMM [6], and the PD method and its deep learning algorithm have been successfully applied to fast MR imaging [9-11]. In the PD process of executing iterative solution with the matrix vector operations, each element in the matrix vector is calculated successively, and the data of one slice is repeatedly calculated until its convergence, according to the principle of iterative algorithm. Then, a loop is added to the outside of the iterative calculation to complete the recon of the whole 3D positive comparison data. The recon time depends on the convergence rate of the iterative algorithm and the scale of the 3D positive contrast data. In these iterative algorithms, the PD method introduces an auxiliary dual variable, which turns the objective function into a saddle point problem. The original variable and dual variable are updated alternately, and its convergence is supported by the strict theory [7, 12]. Based on the quality of the recon image and the speed of convergence as the evaluation criteria, the CPU-based PD method has showed the best performance than others.

The existing CPU-based recon methods perform their calculations sequentially. For image recon tasks, PD method and other iterative processes involve a large number of matrix and vector operations. The sequential execution process based on CPU architecture cannot take advantage of the potential parallelism in matrix vector operation, and the computational efficiency is low. On the other hand, in order to complete the reconstruction of the whole 3D positive contrast image, the CPU-based method needs to set up a cycle to reconstruct one image per cycle, which further increases the time required for reconstruction. These characteristics of CPU-based reconstruction methods lead to the fact that it often takes several minutes for algorithms such as PD to complete the reconstruction of 3D positive contrast images. To raise the

* Research partially supported by the National Natural Science Foundation of China (81729003, 61871373), Natural Science Foundation of Guangdong Province (2018A0303130132), Shenzhen Peacock Plan Team Program (KQTD20180413181834876).

Haifeng Wang and Fang Cai contributed equally to this manuscript.

Haifeng Wang, Fang Cai, Caiyun Shi, Jing Cheng, Zhilang Qiu, Shi Su, and Xin Liu are with Paul C. Lauterbur Research Centre for Biomedical Imaging, Shenzhen Institutes of Advanced Technology, Chinese Academy of Sciences, Shenzhen, Guangdong, China.
  Guoxi Xie is with the Six Affiliated Hospital, Guangzhou Medical University, Qingyuan, China and Department of Biomedical Engineering, Guangzhou Medical University, Guangzhou, China.
  Hanwei Chen is with Department of Radiology, Guangzhou Panyu Central Hospital, China.
  Dong Liang is with Paul C. Lauterbur Research Centre for Biomedical Imaging, Shenzhen Institutes of Advanced Technology, Chinese Academy of Sciences, Shenzhen, Guangdong, China (corresponding author to provide e-mail: dong.liang@ siat.ac.cn).

computational efficiency of the existing CPU-based PD method, we propose to accelerate the PD recon of the positive contrast image by using the multi-core multi-thread features of graphics processing unit (GPU). GPU-based computing takes advantages of the potential parallelism in matrix vector computing and simultaneously rebuilds data on multiple slice layers. Under the experimental condition, that the image quality remains same, the time required for image recon can be significantly reduced.

## II. METHODS

### A. Susceptibility-based positive contrast imaging

The susceptibility-based positive contrast MRI acquires image based on the 3D-SPACE pulse sequence [13-15]. The SPACE sequence can achieve higher signal to noise ratio and use the thinner sections without inter-slice gaps [13]. In addition, it can improve the sample efficiency with isotropy which allows retrospective reformatting to view multiple orientations of the anatomical area being examined. Here, we shifted each readout of the 3D-SPACE sequence by a small amount $T_{shift}$ with a range of 0.2~0.7ms to sample data [14-15]. Finally, two data were acquired (with or w/o echo shift) for measuring the field map induced by the device. After acquiring the 3D data, the susceptibility map is reconstructed by solving a ℓ1 norm optimization problem of the Eq.(1).

$$\chi = \arg\min_{\chi} \frac{\lambda}{2} \| W(D\chi - \Delta B) \|_2^2 + \| MG\chi \|_1 \quad (1)$$

where D is the dipole kernel convolution operator and W is a weighting matrix. G is the gradient operator in all three dimensions. M is the mask matrix. λ is the regularization parameter. ΔB(r) is the local field. To solve the optimization problem of the Eq. (1), the regularized positive contrast inversion problem and its primal-dual formulation is derived. The first-order PD algorithm [8] is applied to solve the $\chi$. The reconstruction framework base on the first-order PD algorithm is suitable for the non-smooth convex optimization problems, and has been proved convergence in mathematics [7]. Then, we can describe the PD algorithm of the susceptibility-based positive contrast MRI as follows [8]:

| **Pseudocode of the PD algorithm in positive contrast MRI** |
|---|
| 1. $L = \|K\|_2, \tau=1/L, \sigma=1/L, \theta = 1, n=0$ |
| 2. Initialize $\chi_0, p_0, q_0$ to zero values |
| 3. $\bar{\chi}_0 = \chi_0$ |
| 4. Repeat<br>  $p_{n+1} = (p_n + \sigma*(W(D\bar{\chi}_n) - \Delta B))/(1+\sigma/\lambda)$    %$prox_\sigma[F_1^*]$<br>  $q_{n+1} = (q_n + \sigma* MG\bar{\chi}_n)/\max(1, \|q_n + \sigma* MG\bar{\chi}_n\|)$  %$prox_\sigma[F_2^*]$<br>  $\chi_{n+1} = \chi_n - \tau(D^*W^*p_{n+1}) - \tau(G^*M^*q_{n+1})$   % $prox_\tau[G]$<br>  $\bar{\chi}_{n+1} = \chi_{n+1} + \theta(\chi_{n+1} - \chi_n)$<br>  n=n+1 |
| 5. Until n≥ N |

Where, $p \in Y, q \in V$ ; $Y$ is the space of field variation; $V$ is the gradient vector space; the whole objective function of our reconstruction problem can be written in the form $F(K\chi)$.

### B. Flow of the proposed scheme

The proposed PD algorithm based on GPUs of the susceptibility-based positive contrast MRI can reserve the image quality and significantly reduce the computational time of the positive contrast MRI reconstruction. Moreover, it can deal with the different sizes of positive contrast MRI data set. The basic steps of the proposed in Figure 1 are as follows:

(1) Preprocess the two data collected by the positive contrast MRI to obtain the dipole kernel convolution operator *D*, the weighting matrix *W*, the mask *Mask*, and the phase contrast field *iFreq_delta*;

(2) Move the data into Step (1) from CPU memory to GPU memory;

(3) Parallelize matrix vector operations by calling multiple GPU threads from the CPU side, based on the multi-core feature of GPU.

(4) Apply the parallelized operations into Step (3) to multiple slice data simultaneously, based on the dynamic parallel features of CUDA (Unified Computing Device Architecture);

(5) Combine Step (3) and Step (4) to implement a parallel iterative calculation method to reconstruct the positive contrast MRI data;

(6) Move the results calculated in Step (5) from CPU memory to GPU memory to display the final reconstructed image of the positive contrast MRI.

### C. GPU Accelerations

Here, three strategies are applied to accelerate the PD reconstruction of the positive contrast MRI, as follows:

(1) Multi-threaded parallel computing strategy

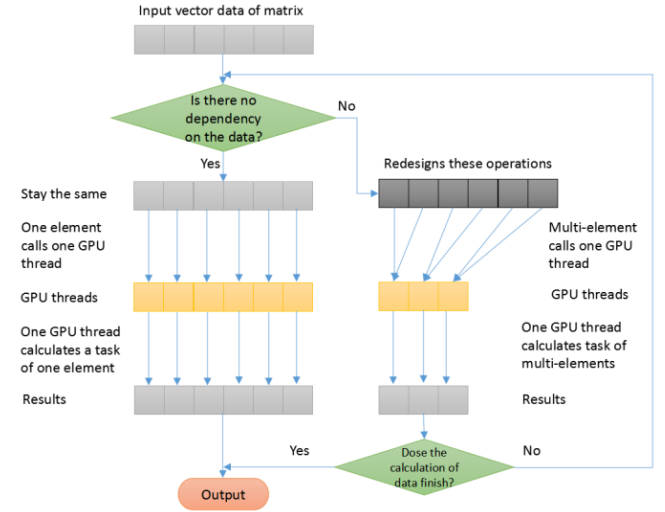

Figure 1. GPU multi-threaded parallel computing module

Modern GPUs generally have thousands of stream processors, and their number of parallel processing tasks far exceeds that of CPUs. The proposed method utilizes the characteristics of the above-mentioned GPU, firstly allocates the required capacity for calculation in the GPU video memory, and then performs the parallelized calculation on the GPU side. Specifically, for each matrix vector operation of the iterative algorithm on the original CPU, first call multiple threads on the GPU side on the CPU side, and then complete the operations executed in a loop on the CPU side on each thread on the GPU side. There is no dependency on the data processed between each thread on the GPU side. Therefore, for data-independent operations, such as matrix transposition and vector addition, the number of GPU-side threads called is the number of elements of the matrix or vector. For data-dependent operations, such as the sum of all elements in a vector, the present method redesigns these operations into a series of small data-dependent operations and then executes them on the GPU. In this case, GPU-side threads are called. The number is often less than the number of elements in the matrix or vector. A matrix multi-threaded parallel computing framework based on GPU is shown in Figure 1.

(2) Multi-data concurrent strategy of dynamic parallelism

The 3D positive contrast MRI requires reconstructing the multi-slice images. Because the tasks of different slices are irrelevant, there is no data dependency among them. The proposed method realizes a concurrent computing framework for multi-layer data in Figure 2. This framework, while retaining the advantages of Strategy (1), uses the concurrent computing strategy of multi-layer data to further utilize the multi-threading GPU features. Specifically, at first, multiple GPU-side threads are started on the CPU side, and the number of started threads is equal to the number of slices, and each thread corresponds to one slice. Then use CUDA dynamic parallel technology to start multiple GPU-side threads from each GPU-side thread to complete the parallelized calculation of the matrix vector described in the Strategy (1). Because CUDA does not provide a fast *Fourier* transform (FFT) static link library on the Windows OS platform, starting multiple GPU-side threads from a GPU-side thread are suitable for positive contrast MRI recon on the Windows OS platform. All matrix vector operations except FFT, and for multi-data FFT calculation, the present method is implemented by setting the number of external loops and adjusting input parameters.

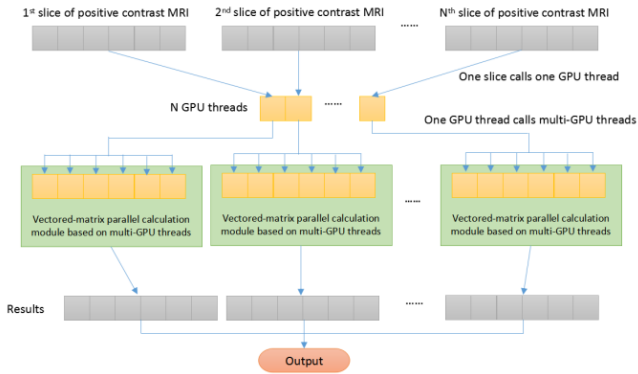

Figure 2. Multi-data concurrent computing module of dynamic parallel

(3) High-efficiency parallelized recon of PD algorithm

In the positive contrast MRI, the CPU-based PD method has shown excellent performance [8]. Based on that algorithm, the proposed method implements an efficient parallelized positive contrast MRI reconstruction, as shown in Figure 3. Specifically, in this framework, the proposed method performs calculations in strict accordance with the iterative steps of the PD method to ensure that the data processing process conforms to the theory of the PD method. Then in each step of the calculation, combine the strategies of Strategy (1) and Strategy (2) to make full use of the GPU resources to achieve efficient calculations. Among them, for simple vectored-matrix calculation, the way of calling GPU threads is relatively simple. In order to save some GPU resources and save the startup of several GPU threads in the middle data slice layer, the proposed method directly calls several times by calling the number of GPU threads to achieve multi-data calculation, and the multiple is the number of layers of positive contrast MRI data. Compared with the CPU-based implementation, the data slice loop and the vectored-matrix operation loop are reduced, which can greatly improve the reconstruction efficiency and reduce the time-cost required.

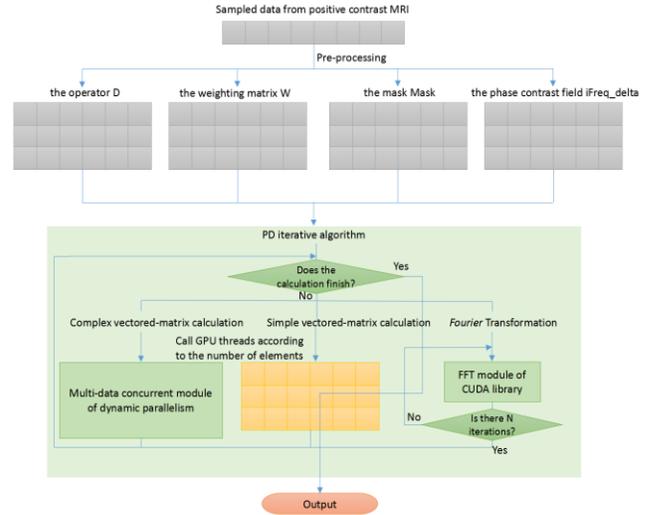

Figure 3. High-efficiency parallelization based on the PD algorithm of the positive contrast MRI. Here, N is less than the number of layers of positive contrast MRI data, and the optimal N is adjusted based on experiments

### D. memory transferring

As stated in the flow of the proposed scheme, the main memory transferring is occurred in the Step (2) and Step (6), which couldn't be avoided. The proposed GPU-based PD reconstruction method minimizes the memory transferring between GPU and CPU using the following two strategies: firstly, all the intermediate data produced by the GPU calculations is stored on the GPU memory, making the next GPU calculation use the intermediate data immediately without data transferring; secondly, the multiple GPU-side threads proposed in the part of "Multi-data concurrent strategy of dynamic parallelism" will control the calculation end points of the multiple slices, and the state variables of the recon process of the multiple slices are stored on the GPU memory, which further decreasing the memory transferring. Therefore, the GPU calculations will not be blocked by the memory transferring, due to these minimization strategies.

## III. EXPERIMENTS

The experimental data were acquired on a 3T MRI scanner (SIEMENS Tim Trio, Erlgen, Germany) [8,14-16]. The dataset were acquired using three metal devices. The 3D data of the proposed method were acquired with or without $T_{shift}$ of 0.6 ms. And the experiments were carried out on the positive contrast MRI of three metal devices with different data sizes: the stent (Figure 4), the seed embedded in the meat (Figure 5) and the needle (Figure 6). Relevant imaging parameters were as follows: TE/ TR: 33/1500ms, in-plane resolution: [0.72, 0.72, 2] $mm^3$; BW: 698Hz/pixel. The experimental results showed that the speed of the GPU reconstruction calculation is 4~15 times than the CPU-based solution.

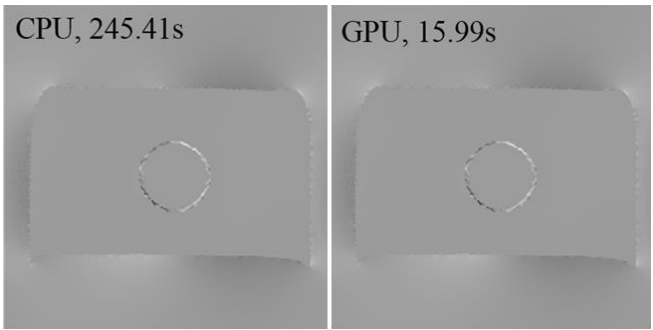

Figure 4. The cross-section reconstruction images of the stent from the positive contrast MRI

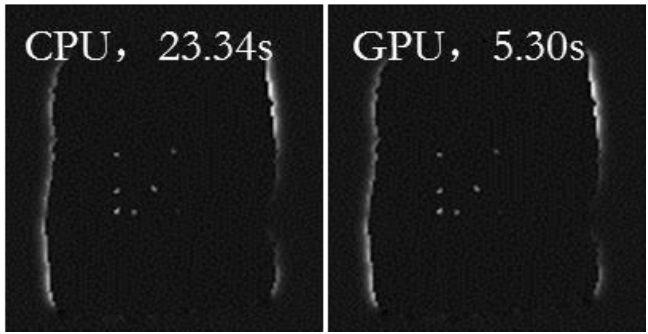

Figure 5. The reconstruction results of seeds embedded into meat from the positive contrast MRI reconstruction

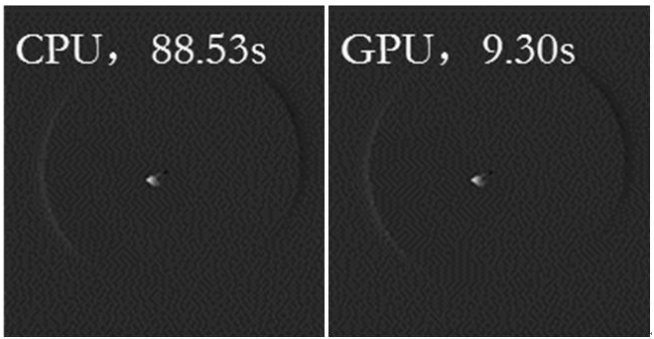

Figure 6. The PD reconstruction results of one needle in a water phantom from the positive contrast MRI

Figure 4 was the contrast results of stent. The data size was 256×258×25. The left was from the CPU-based method, and the recon time was 245.41 seconds. The right was from the proposed GPU-based method, and the recon time was 15.99 seconds. The image qualities of them were almost same. The GPU-based speed was 15 times faster than CPU. Figure 5 was the results of the seeds embedded in the meat. The data size was 128×132×12. The left was the result of the CPU-based method, the recon time was 23.41s seconds. The right was the result of the proposed method, and the recon time was 5.3 seconds. The images had same qualities. The proposed method was 4 times faster than the former. Figure 6 was from the needle which was inserted into a water phantom doped with 1.0 g/L copper sulfate solution. The data size was 128 × 132 × 42. The left was from the CPU-based method, and the recon time was 88.53 seconds. The right was from the proposed GPU-based recon method, and the recon time was 9.3 seconds. The images reconstructed by the two methods were almost same. The recon speed of the proposed method was 9 times faster than the CPU-based method.

## IV. CONCLUSION

The existing CPU-based iterative recon method of the positive contrast MRI does not take advantage of the potential parallelism of vectored-matrix operations. It is necessary to set a loop to operate on the elements in the vectored-matrix in turn inside the loop. Here, we proposed the GPU-based PD recon method of the positive contrast MRI to makes full use of the potential parallelism of vectored-matrix operations. This proposed method of the positive contrast MRI can directly reduce the one-layer loop of the former CPU-based method under any circumstances, further enhancing the practicability of the positive contrast MRI in the clinical applications.